\documentclass[aps,apl,twocolumn,showpacs,superscriptaddress]{revtex4}

\usepackage{graphicx} 
\usepackage{soul}
\usepackage{multirow}
\usepackage[T1]{fontenc}
\usepackage[latin9]{inputenc}
\usepackage{rotating}
\usepackage{amsmath}
\usepackage{amssymb}
\usepackage{tabularx}

\begin{document}

\title{Graphane Nanoribbons: A Theoretical Study}

\author{H. \c{S}ahin}
\affiliation{UNAM-Institute of Materials Science and
Nanotechnology, Bilkent University, 06800 Ankara, Turkey}

\author{C. Ataca}
\affiliation{UNAM-Institute of Materials Science and
Nanotechnology, Bilkent University, 06800 Ankara, Turkey}
\affiliation{Department of Physics, Bilkent University, 06800 Ankara,
Turkey}

\author{S. Ciraci}\email{ciraci@fen.bilkent.edu.tr}
\affiliation{UNAM-Institute of Materials Science and
Nanotechnology, Bilkent University, 06800 Ankara, Turkey}
\affiliation{Department of Physics, Bilkent University, 06800 Ankara,
Turkey}

\date{\today}

\begin{abstract}
In this study, we investigate the electronic and magnetic properties
of graphane nanoribbons. We find that zigzag and armchair graphane
nanoribbons with H-passivated edges are nonmagnetic semiconductors.
While bare armchair nanoribbons are also nonmagnetic, adjacent
dangling bonds of bare zigzag nanoribbons have antiferromagnetic
ordering at the same edge. Band gaps of the H-passivated zigzag and
armchair nanoribbons exponentially depend on their width. Detailed
analysis of adsorption of C, O, Si, Ti, V, Fe, Ge and Pt atoms on
the graphane ribbon surface reveal that functionalization of
graphane nanoribbons is possible via these adatoms. It is found that
C, O, V and Pt atoms have tendency to replace H atoms of graphane.
We showed that significant spin polarizations in graphane can be
achieved through creation of domains of H-vacancies and
CH-divacancies.

\end{abstract}


\maketitle

\section{Introduction}\label{Introduction}

After the synthesis of two dimensional (2D)
graphene\cite{novo1,novo2,zhang,berger,kats,geim}, its
nanoribbons (NRs) have been a subject of interest, both
experimentally\cite{han,li} and
theoretically.\cite{lee,barone,abanin,cresti,ezawa0,gunlycke2,hod1} Advances in experimental techniques are paving the way for
integrating the exceptional electrical, optical and magnetic
functionalities of these nanometer-sized materials into future
electronic technology. In addition to numerous theoretical and experimental works on
graphene and its NRs, research efforts have been also
devoted to synthesize its various derivatives. Notably,
the synthesis of a 2D hydrocarbon in honeycomb
structure,\cite{elias} namely graphane, followed its prediction through theoretical works.\cite{sofo,boukhvalov}
Detailed analysis of hydrogenation processes of graphene leading
to graphane and the existence of hydrogen (H) frustrations were discussed very
recently.\cite{Graphene to graphane} Furthermore, it was reported
that graphane NRs have more favorable formation
energies than experimentally available graphene ribbons.
\cite{yafei} Recently, we reported the possibility of obtaining
magnetization through dehydrogenation of domains on 2D
graphane.\cite{hasan-can} Modification of electronic structure of graphane by introducing either a hydroxyl group or a H-vacancy
was also investigated with GW self-energy calculations.
\cite{lebeque} Although there are several recent studies on
graphane, electrical and magnetic properties of its NRs
have remained unexplored.

Recent advances in graphane have motivated us to explore the
properties of zigzag and armchair graphane NRs. In this
paper, using first-principles plane wave method within the
density functional theory (DFT) we investigated the electronic
and magnetic properties of bare and H- passivated graphane
NRs. We also explored the effects of specific
imperfections on these properties. These are various foreign
atoms adsorbed on the surfaces of graphane NRs, vacancies and
edge roughness. We found that these imperfections can attribute
interesting functionalities by modifying the electronic and
magnetic properties of graphane NRs.

\begin{figure}
\includegraphics[width=8.5cm]{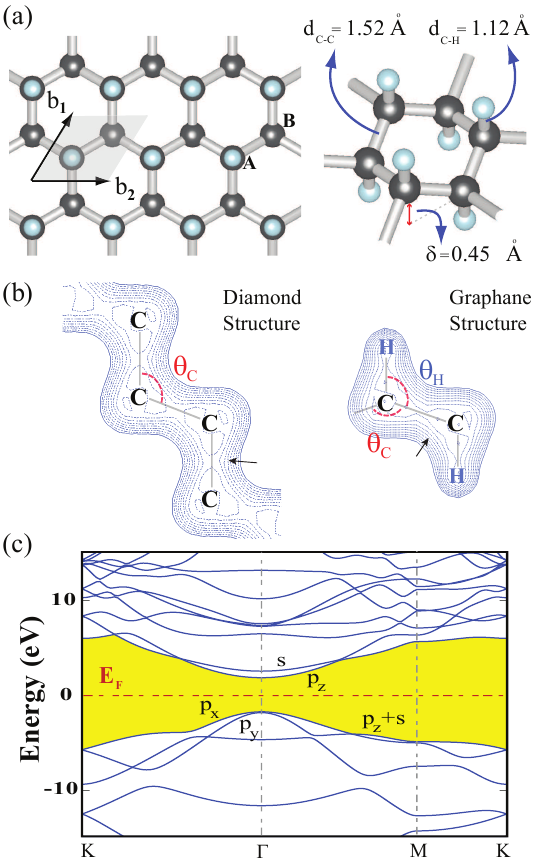}
\caption{(Color online) (a) Top and perspective view of the atomic
structure of infinite 2D graphane sheet having honeycomb structure.
Two sublattices of graphane are indicated by A and B. Black (dark)
and blue (light) balls are for carbon and hydrogen atoms,
respectively. (b) Charge density contour plots of diamond and
graphane are shown on a plane passing through C-C-C-C and H-C-C-H
bonds, respectively. The tetrahedral angle of the diamond $\theta_C
= 109.47^o$. Arrows indicate the direction of increasing charge
density. The calculated values of $\theta_C$ and $\theta_H$, namely
C-C-C and H-C-C bond angles in graphane respectively, are given in
Table~\ref{tab:graphene-graphane}. Contour spacings are 0.0286
e/\AA$^3$. (c) The LDA energy band structure where the orbital
character of specific bands is also given. The band gap is shaded
yellow/gray.} \label{2d-graphane}
\end{figure}

\section{Methods}\label{Methods}

We perform first-principles calculations\cite{kresse} within DFT using
projector augmented wave (PAW) potentials\cite{paw} and
approximate exchange-correlation functional by spin polarized
local density approximation\cite{lda} (LDA). Kinetic energy
cutoff $ \hbar^2 |\mathbf{k}+\mathbf{G}|^2 / 2m $ for plane-wave
basis set is taken as 500 eV. In the
self-consistent potential and total energy calculations of 2D graphane a set
of (35x35x1) \textbf{k}-point sampling is used for Brillouin
zone (BZ) integration. The convergence
criterion of self consistent calculations for ionic relaxations
is $10^{-5}$ eV between two consecutive steps. By using the
conjugate gradient method,
all atomic positions and unit cells are optimized until the
atomic forces are less than 0.03 eV/\AA. Pressures on the
lattice unit cell are decreased to values less than 1.0 kBar.

In order to correct the energy bands and band gap values obtained
by LDA, frequency-dependent GW$_0$  calculations are carried
out.\cite{gw} Screened Coulomb potential W, is kept fixed
at initial DFT value W$_0$ and Green's function G, is
iterated five times. GW$_0$ self-energy calculations
are carried out with $15 $ \AA ~ vacuum  spacing,
default cut-off potential, 160 bands and 64 grid points.

Graphane NRs are treated by the supercell geometry
within periodic boundary conditions. Vacuum spacing of at least $15 $ \AA~ is placed between adjacent graphane NRs to hinder the  interactions. In specific cases double unit cell is
used in our calculations to allow the possible
antiferromagnetic (AFM) orderings along the ribbon edges.
Reciprocal space integrations are carried out with 1x1x15
Monkhorst-Pack \textbf{k}-point grids.

\section{Two Dimensional Graphane}\label{2D}

For a better understanding of graphane NRs we start with 2D graphane
which is derived from the hydrogenation of graphene, where each
carbon atom is saturated by a single hydrogen atom. Accordingly, the
primitive cell of graphane consists of two carbon and two hydrogen
atoms. Chair like configuration of infinite sheet of 2D graphane is
formed, whereby each carbon atoms of A- and B-type sublattices are
saturated by hydrogen atoms from above and below, respectively, as
described in Fig.~\ref{2d-graphane}(a). This is known as the most
favorable and stable hydrocarbon conformation.\cite{sofo, hasan-can}
The planar graphene honeycomb structure, which is stabilized by
planar $sp^2$-hybrid orbitals and $\pi$-bonding between adjacent
perpendicular $p_z$-orbitals is puckered (buckled) as a result of
the adsorption of H atoms, whereby a single C atom is bound to three
adjacent C atoms and a single H atom through 'tetrahedrally
coordinated' $sp^{3}$-like hybrid orbitals. The buckling, $\delta$,
i.e. the perpendicular distance between A-type and B-type carbon
sublattices, is calculated to be 0.45 \AA. As a result, the lattice
constant of graphene increases from 2.46~\AA~ to 2.51~\AA~ and hence
the C-C and C-H bond lengths become 1.52 and 1.12~\AA,
respectively, as indicated in Fig.~\ref{2d-graphane}(a). The charge
density contour plots given in Fig.~\ref{2d-graphane}(b) indicate
that the charge distribution in graphane especially around the C-C
bond is reminiscent of that in tetrahedrally coordinated diamond. In
graphane, the angle $\theta_H$ between H-C and C-C bonds and
$\theta_C$ between adjacent C-C bonds are $107.35^o$ and $111.51^o$,
respectively. The mean value of these angles is equal to the
tetrahedral angle $\theta_C$ of diamond. The maximum of the C-C bond
charge in diamond and in graphane are found to be $\sim 0.282$ and
$\sim 0.286$ e/\AA$^3$, respectively.

\begin{table*}
\caption{Comparison of the calculated quantities of graphene and
graphane. Lattice constant, $a$; C-C bond distance, $d_{C-C}$; C-H
bond distance, $d_{C-H}$; the buckling, $\delta$ [see
Fig.~\ref{2d-graphane}(a)]; angle between adjacent C-C bonds,
$\theta_{C}$ [see Fig.~\ref{2d-graphane}(b)]; angle between adjacent
C-H and C-C bonds, $\theta_{H}$;  energy band gap calculated by LDA,
$E_{g}^{LDA}$; energy band gap corrected by GW$_0$, $E_{g}^{GW_0}$;
cohesive energy $E_{c}^{nm}$, ($E_{c}^{m}$) obtained with respect to
nonmagnetic (magnetic) free atom energies; the C-H bond energy,
$E_{C-H}^{nm}$ ($E_{C-H}^{m}$) obtained with respect to nonmagnetic
(magnetic) free atom energies; photoelectric threshold (work
function), $\Phi$; in-plane stiffness, $C$ and Poison ratio, $\nu$.
} \label{tab:graphene-graphane}
\begin{center}
\begin{tabular}{cccccccccccccccc}
\hline  \hline
Material &       $a$    & $d_{C-C}$  & $d_{C-H}$ & $\delta$ & $\theta_{C}$ & $\theta_{H}$ &$E_{g}^{LDA}$ &  $E_{g}^{GW_0}$  &    $E_{c}^{nm}$    &   $E_{c}^{m}$    &     $E_{C-H}^{nm}$     & $E_{C-H}^{m}$     & $\Phi$     & $C$     & $\nu$\\
(1x1) unit cell &(\AA)     & (\AA)   & (\AA)   &    (\AA) &  (deg) & (deg) &($eV$)  &  ($eV$)   &     ($eV$)           &  ($eV$)             &      ( $eV$ )          &    ($eV$)             &   ($eV$)    &  ($J/m^{2}$)   &\\
\hline \hline
Graphene & 2.46 & 1.42 & - & - & 120 & - &  0.00 & 0.00 &20.16 & 17.87& -& - &4.77 & 335\cite{lee-science} &0.16\\
\hline
Graphane & 2.51 & 1.52& 1.12 & 0.45 & 111.51 &107.35 &3.42& 5.97 & 27.65 & 23.57 &3.74 &2.84 & 4.97& 243\cite{topsakal-graphane} &0.07\\
\hline  \hline
\end{tabular}
\end{center}
\end{table*}

The cohesive energy of graphane (per unit cell) relative to free C
and H atoms is obtained from $E_c = 2\times E_T^H+2\times
E_T^C-E_T^{Gra}$ , where $E_T^H$ and $E_T^C$ are the total energies
of single free C, and free H, whereas $E_T^{Gra}$ is the total
energy of graphane. The cohesive energy depends whether the energies
of magnetic or nonmagnetic (NM) states of free C and free H atoms
are considered. Here the cohesive energy per unit cell of graphane
is calculated to be 23.57 (27.65) eV by considering the magnetic
(nonmagnetic) states of free atoms. As for the average C-H bond
energy, one can use the formula $E_{C-H}=
(-E_T^{Gra}+E_T^{Gre}+2\times E_T^H)/2$ where $E_T^{Gre}$ is the
total energy of graphene. The calculated value is 2.84 (3.74 for
nonmagnetic case). We reported\cite{hasan-can} that the desorption
of a single H atom from the graphane is an endothermic reaction with
an energy of 4.79 eV. When compared with the energy of C-H bond, the
desorption energy of single H is larger, since the single H removal
creates reconstruction of nearby atoms which further reduces the
total energy of the system.

Due to $sp^{3}$-saturation of C atoms, 2D graphane is a NM
semiconductor with a direct band gap of 3.42 eV as shown in
Fig.~\ref{2d-graphane}(c). However, this band gap, which is
underestimated within DFT, is corrected by GW$_{0}$ approximation to
become 5.97 eV.\cite{hasan-can, lebeque} While the top of the
valence band originates from $p_{x}$- and $p_{y}$-orbitals of C
atoms, the bottom of the conduction band has C-$p_{z}$ orbital
character. Earlier, we also calculate\cite{hasan-can} the phonon
modes of infinite 2D graphane, which yield all real frequencies in
BZ. Having all frequencies positive indicate the stability of
graphane structure.

An exceptional feature of graphane is related with its
interesting charging configuration. While graphene has a
covalent bonding, upon the saturation of each C atom by a
single H atom, a charge of $\delta Q\sim$ 0.1 electrons is
transferred from H to C. At the end, the negatively charged
bilayer of carbon atoms becomes sandwiched between positively
charged H layers. Graphane having this charging has the photoelectric threshold (work
function) calculated to be $\Phi$=4.97 eV, which is 0.2 eV
larger than that of graphene. In Table~\ref{tab:graphene-graphane}, we presented the calculated values related with structure, energy bands, photoelectric threshold etc. of graphane and graphene for the sake of comparison. To be complete Table~\ref{tab:graphene-graphane} also includes calculated elastic constants such as in-plane stiffness\cite{brandbyge} $C ~[=\frac{1}{A}( \frac{\partial^2 E_T}{\partial \epsilon^2})$, where $E_T$ is the total energy of the system under strain, $\epsilon$ is the uniaxial strain and $A$ is the area of the unit cell] and Poisson's ratio\cite{hasan-ansiklopedi} $\nu$.

\section{Graphane Nanoribbons}\label{Graphane NRs}

\begin{figure}
\includegraphics[width=8.5cm]{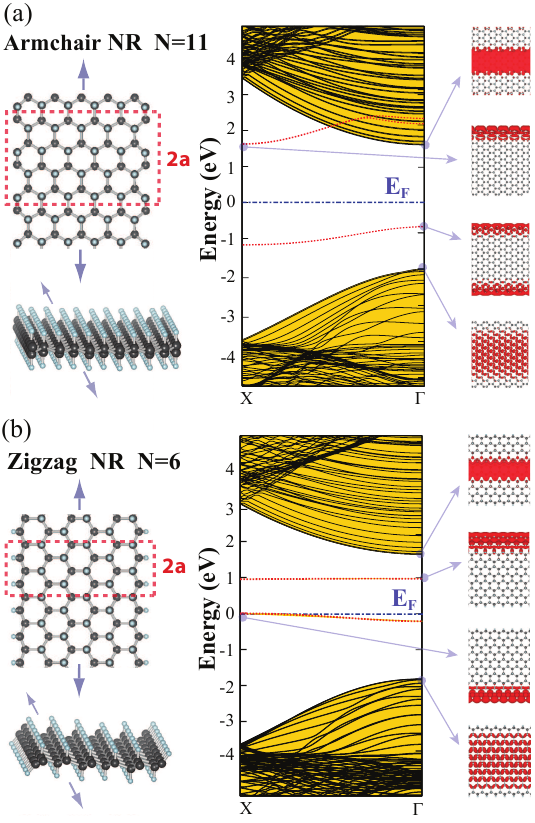}
\caption{(Color online) (a) Atomic structure of bare armchair
graphane NR having $N$=11. The double unit cell of the nanoribbon is
delineated by red/dashed lines with the lattice constant $2a$.
Large/black and small/light blue balls indicate carbon and hydrogen
atoms. Energy band structure corresponding to the armchair graphane
NR and charge density of selected bands are shown in the panels at
the righthand side. (b) Atomic structure of bare zigzag graphane NR
having $N$=6 with double unit cell delineated by red/dashed lines
and with lattice constant $2a$. Energy band structure and isosurface
charge density of selected states corresponding to zigzag graphane
NR are indicated. Bands shown by red/dotted lines are derived from
edge states. Zero of energy is set to the Fermi level, shown by
dash-dotted lines, of the nanoribbons with H-passivated edges. In
spin polarized calculations double unit cell is used to allow
antiferromagnetic order along the edges.} \label{ribbon1}
\end{figure}

The nanoribbons cut from a 2D graphane are structures providing
important features for various technological applications. Two major
families of graphane NRs are distinguished depending on their
orientations; namely armchair, and zigzag graphane NRs. Apart from
the orientation, graphane NRs are characterized by their widths. For
armchair graphane NRs, the width, $N$, is defined by the number of
C-C dimers in the unit cell which are parallel to the axis of the
nanoribbon; for zigzag graphane NRs, $N$ denotes the number of
zigzag carbon chains along the nanoribbon axis. The electronic
structure changes depending on whether the dangling bonds of carbon
atoms at the edge are free (bare) or passivated by single H atoms.
In Fig.~\ref{ribbon1}(a) and ~\ref{ribbon1}(b) we show the atomic
and electronic structure of armchair and zigzag graphane NRs with
$N$=11 and $N$=6, respectively.

Bare armchair graphane NR is a NM indirect band gap semiconductor,
since the electrons at the dangling orbitals are paired. The band
gap of 2D graphane is reduced due to the bands of edge states of
dangling bonds occurring in the band gap. The charge density
analysis of these edge state bands presented in
Fig.~\ref{ribbon1}(a) clearly shows that they are localized at the
edges of the NR. Upon hydrogen passivation of the threefold
coordinated edge atoms, the edge state bands in the gap disappear
and the direct band gap opens up.

Similar to armchair graphane NRs, the dangling bonds of the
threefold coordinated carbon atoms at both edges of zigzag graphane
NRs give rise to edge states in the band gap. Owing to relatively
large distance between adjacent dangling bonds and hence their
relatively smaller coupling, they have a minute dispersion. Charge
density isosurfaces of these edge states confirm their localization
at the edges. Also, owing to their weak coupling the spins of
adjacent dangling bonds cannot be paired. Consequently, each carbon
atom at the zigzag edge attains a magnetic moment of 1  Bohr
magneton ($\mu_B$). The ordering of these magnetic moments is
however dramatically different from that of the zigzag graphene NRs.
The significant coupling between two spin states at different edges
gives rise to AFM ordering along the graphene edge.

\begin{figure}
\includegraphics[width=8.5cm]{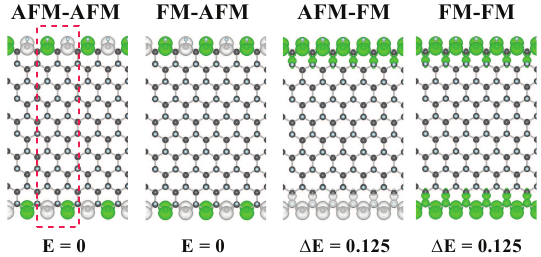}
\caption{(Color online) Total energies of possible magnetic
orderings at the edges of bare zigzag graphane nanoribbons.
Calculations are performed in double unit cell delineated by
red/dashed lines. Spin up and spin down states are shown by
green/dark and grey/light isosurfaces of the difference charge
density, $\Delta \rho$.}

\label{ribbon2}
\end{figure}

The physical origin of the magnetic interactions between the edges
of graphitic fragments was treated earlier.\cite{edge-states} For
zigzag graphene nanoribbon, collective alignment of the magnetic
moments along the same ribbon edge through the ferromagnetic (FM)
interaction, but AFM ordering between opposite edges is attributed
to magnetic tail interactions. The magnetic ordering of a wide
zigzag graphane NRs ($\sim$ 2 nm) are examined in a supercell
comprising two unit cells, whereby antiferromagnetic ordering
between adjacent dangling bonds at the same edge is allowed. We
consider four different orderings, such as AFM-AFM (where,
respectively, the ordering of magnetic moments at the atoms located
at different and same edge are antiferromagnetic), FM-AFM, AFM-FM
and FM-FM as described in Fig.~\ref{ribbon2}. Here the difference
charge density is defined as the $\Delta
\rho=\rho^{(\uparrow)}-\rho^{(\downarrow)}$ where $\rho^{\uparrow
(\downarrow )}$  is spin up (down) charge. Since the magnetic
interaction between two edges vanishes, AFM-AFM and FM-AFM
orderings have the same total energy corresponding to the ground
state. These orderings are 125 meV energetically more favorable than
the AFM-FM and FM-FM orderings. From this analysis it is revealed
that AFM spin alignment between adjacent atoms at the same edge is
preferred by the zigzag graphane NRs. FM ordering between two spins
has 62.5 meV higher energy. For zigzag graphane NR with $N$=6, the
magnetic coupling between two edges is negligible and hence either
AFM-AFM or FM-AFM orderings have the same energy and they are ground
states. However, when the width of the nanoribbon is less than
12~\AA~the degeneracy of AFM-AFM and FM-AFM states is broken. The
bare zigzag graphane NR is an indirect and antiferromagnetic
semiconductor with an indirect band gap relatively smaller than that
of 2D graphane. However, it becomes NM, direct band gap
semiconductor upon passivation of the dangling bonds. Also the
magnetic edge states disappear and band gap becomes larger.

\begin{figure}
\includegraphics[width=8.5cm]{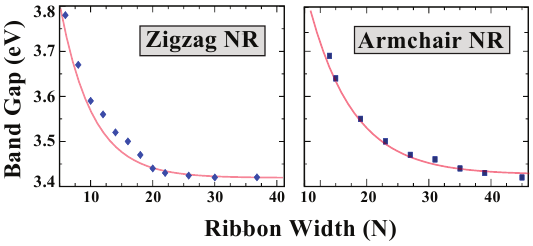}
\caption{(Color online) Variation of the energy band gap of
H-passivated zigzag and armchair NRs of graphane as a function of
width given by $N$. The variation of the band gap with $N$ is fitted
to the curve given by continuous line. (See text)}

\label{ribbon3}
\end{figure}

The value of the band gap of graphane NRs can differ from that of parent 2D structure due to a combined effects of quantum confinement, edge state bands as well as folding in the direction perpendicular to the NR axis. We note that the band gap of graphane NR depends on its width given by N. However, the quantum confinement effect is not as emphasized as that found in graphene NR. In Fig.~\ref{ribbon3}, H-passivated armchair and zigzag graphane NRs display a quantum confinement effect, namely the band gap reduction with increasing width or $N$. This behavior is fitted to an expression as:

\begin{equation}
E_{gap}(N)=3.42~eV +\alpha\exp(-N\beta)
\end{equation}
Here $\alpha$ and $\beta$ are fitting parameters. The values of
$\alpha$ and $\beta$ are found to be 1.18 eV and 0.19 for zigzag
graphane NRs (2.15 eV and 0.14 for armchair graphane NRs). The band
gaps of both types of graphane NRs go to 3.42 eV as $N \rightarrow$
$\infty$. Considering the GW$_0$ corrected value of the band gap,
5.97 eV, the above fitting can be expressed as
$E_{gap}(N)=5.97+\alpha\exp(-N\beta)$ assuming that the same scissor
operation is applicable for graphane NRs. Apparently, the band gap
values of graphane NRs calculated using DFT are underestimated and
hence their values are expected to occur $\sim$ 2.5 eV larger than
presented in Fig.~\ref{ribbon3}. Unfortunately, GW$_{0}$ self-energy
corrections cannot be performed for graphane NRs due to large
computational time. We note that the variation of band gap is
relevant for $N<22$ for zigzag graphane NRs and $N<30$ for armchair
graphane NRs. In Ref.[\onlinecite{yafei}] the band gaps of hydrogen
saturated graphane NRs were calculated for $6 \leq N \leq 15$ using
Generalized Gradient Approximation (GGA). In the present study band
gaps are calculated for $10 \leq N \leq 40$ using LDA. The
hydrogenated graphane NRs included in both studies, namely those $10
\leq N \leq 15$, are in good agreement. Largest band gap difference
of 0.1 eV found between two studies is attributed to different
approximations (i.e LDA versus GGA) in representing
exchange-correlation potential.

\section{Functionalization of Graphane NRs by Adatoms}\label{Adatoms}

\begin{figure}
\includegraphics[width=8.5cm]{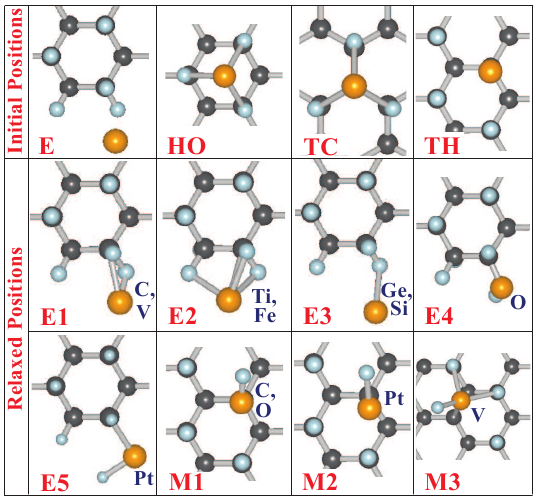}
\caption{(Color online) Schematic representations of possible
positions of adatoms on a large H-passivated armchair graphane
nanoribbon. Positions of adatoms obtained after the structure
optimization through minimization of total energy and forces
exerting on the atoms are also shown.}

\label{adatom}
\end{figure}

Adsorption of adatoms is widely used and an efficient way to provide
new functionalities to structures in nanoscale
applications.\cite{hasan-prb,can1,cohen-adsorp} Adsorption of C, O,
Si, Ge, Pt, V, Fe and Ti adatoms on graphane nanoribbons is
investigated by using (1x1x6) supercell of H-passivated armchair
graphane NR having $N$=8. Adsorption geometry is determined by
calculating the lowest binding energy corresponding to the optimized
structures. To this end, the binding energies of adatoms at four
initial sites described in Fig.~\ref{adatom}: These are edge site E;
hollow site HO, top site of the carbon TC and top site of the
hydrogen atom TH in the middle of NR. The stable binding sites are
determined upon structure relaxation. In the same figure the
positions of adatoms after relaxation are also shown. Binding
energies ($E_b$) of these atoms are calculated by using the
expression
\begin{equation}\label{equ:binding}
E_b={-E_{T}[NR+A]} + E_{T}[NR] + E_{T}[A]
\end{equation}
in terms of the total energies of the optimized structure of
graphane NR with adatom, $E_{ T}[NR+A]$; bare NR, $E_{T}[NR]$ and
free adatom, $E_{T}[A]$. Owing to the large graphane nanoribbon unit
cell, \textbf{k}-point sampling is done only along the axis of the
ribbon. All the energies are calculated in the same unit cell and
obtained from the lowest ground-state total energies (either
magnetic or nonmagnetic). Initially, at least 2~\AA~ distance
between adatom and outermost graphane atom is provided to find the
relaxation site of adatom on the ribbon.

An atom adsorbed on graphane NR may give rise to resonance states in
the valence and conduction bands, and also localized states in the
band gap. Owing to the periodic boundary conditions, the energy
levels associated with these adatoms may form energy bands. Since we
consider a supercell consisting of six unit cell of H-passivated
armchair graphane NR, the adatom-adatom distance is large and
prevents coupling between adjacent adatoms. In this respect, the
flat bands associated with adatoms can mimic either resonance or
localized states relevant for dilute doping.

In Table~\ref{tab:adsorp}, we present nearest neighbor distances
between adatom and H ($d_{H}$) and C ($d_{C}$), binding energy
($E_{b}$), total magnetic moment of the system ($\mu_{T}$), charge
transfer between adatom and adsorbate ($\rho*$). Since the inversion
symmetry of graphane can be broken by the adsorption of an atom, a
net electric-dipole moment can be induced thereof. Thus electric
dipole moment values calculated in the direction normal to graphane
NR surface p, induced by adatom are also listed in
Table~\ref{tab:adsorp}. Photoelectric threshold of the
adatom+graphane system perpendicular to the plane of NR, $\Phi$, is
calculated by the difference of the electrostatic potential at
distances where the gradient of it is negligible and the Fermi level
of the adsorbed system. It should be noted that both p and $\Phi$ is
relevant only for uniform coverage of graphane. For a dilute
impurity, these calculated values converge to that of armchair
graphane. When the width of armchair graphane NR is large enough,
one expects the photoelectric threshold, $\Phi$, and dipole moment,
p, to converge to 4.97 eV and 0 e\AA, respectively.\cite{hasan-can}

\begin{table*}
\caption{Summary of the calculated quantities for adatoms adsorbed
on a H-passivated armchair graphane NR. The first and second lines
in each row associated with a given adatom adsorbed to edge site and
the sites near the center of the graphane NRs, respectively. $d_H$,
the adatom-nearest hydrogen distance; $d_C$, the nearest adatom
carbon distance; $E_b$, adatom binding energy; $\mu_{T}$, magnetic
moment per supercell; $\rho^{*}$, excess charge on the adatom (where
negative sign indicates excess electrons); $\Phi$, photoelectric
threshold (work function); p, dipole moment calculated in the
direction normal to graphane NR surface; $E_i$ energies of localized
states associated with adatoms. Localized states are measured from
the top of the valence bands in eV. The occupied ones are indicated
by bold numerals and their spin alignments are denoted by either
$\uparrow$ or $\downarrow$. Up to first seven adatom states of
$E_{i}$ are shown. } \label{tab:adsorp}
\begin{center}
\begin{tabular}{ccccccccccc}
\hline  \hline
$Atom$&$Site$ &$d_H$ & $d_C$ & $E_b$& $\mu_{T}$ & $\rho^{*}$ & $\Phi$& p & $E_{i}$  &\\
&&(\AA) & (\AA) & ($eV$)& $(\mu_{B})$ & (e)& $(eV)$& (e$\times$\AA)& $\uparrow$: Spin-up,  $\downarrow$: Spin-down States&\\
\hline \hline
\textbf{C}  & $E \rightarrow E1$   & 1.27 & 2.05 & 1.10 & 2.00& -0.28& 3.85&0.02& \textbf{0.45$\uparrow$}, \textbf{0.88$\uparrow$}, 2.17$\downarrow$, 2.88$\downarrow$, 3.53$\uparrow$,   3.64$\downarrow$ \\
\textbf{}   & $TH \rightarrow M1$  & 1.12 & 1.45 & 3.99 & 0.00& -0.18& 4.02& 0.34 &\textbf{0.72, 1.52}  \\

\hline
\textbf{O}  & $E \rightarrow E4$   & 0.98 & 1.42 & 5.67 & 0.00& -1.23& 4.63&0.19& \textbf{-0.87, -0.50, -0.22, 0.00}, 4.80  \\
\textbf{}   & $TH \rightarrow M1$  & 0.98 & 1.43 & 5.89 & 0.00& -1.27& 4.59& 0.11& \textbf{-1.16, -1.02, -0.66}, 4.08 \\

\hline
\textbf{Si}  & $E \rightarrow E3$   & 1.88 & 2.65 & 0.54 & 2.00& 0.03& 3.42&0.03& \textbf{1.20$\uparrow$, 1.34$\uparrow$}, 2.36$\uparrow$, 2.43$\downarrow$, 2.56$\downarrow$, 2.95$\downarrow$\\
\textbf{}    & $TH \rightarrow TH$  & 1.84 & 3.00 & 0.68 & 2.00& -0.02& 3.59& 0.42& \textbf{1.23$\uparrow$, 1.24$\uparrow$}, 1.93$\uparrow$, 2.42$\downarrow$, 2.45$\downarrow$, 2.46$\downarrow$ \\

\hline
\textbf{Ti}  & $E \rightarrow E2$   & 1.97 & 2.16 & 0.81 & 2.00& 0.19& 2.65&0.27& \textbf{1.77$\uparrow$, \textbf{2.30$\downarrow$}, 2.34$\uparrow$, 2.43$\uparrow$}, 2.81$\uparrow$, 2.89$\uparrow$, 3.11$\uparrow$\\
\textbf{}    & $TC \rightarrow TC$  & 1.95 & 2.52 & 0.89 & 3.91& 0.31& 2.16& 0.92& \textbf{2.18$\uparrow$, 2.48$\uparrow$, 2.49$\uparrow$, 2.95$\uparrow$}, 3.18$\downarrow$, 3.27$\uparrow$, 3.28$\uparrow$ \\

\hline
\textbf{V}  & $E \rightarrow E1$   & 1.99 & 2.40 & 0.54 & 5.00& 0.14& 2.56&0.22& \textbf{1.49$\uparrow$, 2.10$\uparrow$, 2.38$\uparrow$, 2.43$\uparrow$, 2.50$\uparrow$}, 2.70$\uparrow$, 3.16$\downarrow$\\
\textbf{}    & $TC \rightarrow M3$  & 1.71 & 2.07 & 1.24 & 3.00& 0.93& 3.13& 0.43& \textbf{0.10$\uparrow$, 0.49$\downarrow$, 1.45$\uparrow$, 1.60$\uparrow$, 1.84$\uparrow$}, 2.20$\uparrow$,  3.03$\downarrow$\\

\hline
\textbf{Fe}  & $E \rightarrow E2$   & 1.90 & 2.15 & 0.57 & 4.00& 0.05& 3.21&0.36& \textbf{-0.42$\uparrow$, -0.27$\uparrow$, -0.08$\uparrow$, -0.03$\uparrow$, 0.03$\uparrow$}, 1.48$\downarrow$, 1.59$\uparrow$ \\
\textbf{}   & $HO \rightarrow HO$  & 1.95 & 2.83 & 0.65 & 4.00& 0.07& 2.98& 0.53& \textbf{\textbf{-1.15$\downarrow$}, -0.75$\uparrow$, -0.57$\uparrow$, -0.41$\uparrow$, -0.34$\uparrow$, 0.09$\uparrow$ }, 1.39$\uparrow$\\

\hline
\textbf{Ge}  & $E \rightarrow E3$   & 1.96 & 2.71 & 0.48 & 2.00& 0.00& 3.38&0.03& \textbf{1.27$\uparrow$, 1.38$\uparrow$}, 2.25$\uparrow$, 2.41$\downarrow$, 2.52$\downarrow$, 2.81$\downarrow$\\
\textbf{}    & $TH \rightarrow TH$  & 1.91 & 3.06 & 0.62 & 2.00& -0.02&3.55 & 0.36& \textbf{1.30$\uparrow$, 1.31$\uparrow$}, 1.89$\uparrow$, 2.38$\uparrow$, 2.42$\uparrow$, 2.43$\uparrow$ \\

\hline
\textbf{Pt}  & $E \rightarrow E5$   & 1.56 & 2.00 & 3.79 & 0.00& -0.13& 4.57&-0.09& \textbf{-0.78, -0.36, -0.13}, 0.13, 2.76 \\
\textbf{}   & $TH \rightarrow M2$  & 1.54 & 2.04 & 3.50 & 0.00& -0.18& 4.30& 0.34&  \textbf{-0.36, -0.21, 0.38}, 2.58, 3.43 \\

\hline \hline

\end{tabular}
\end{center}
\end{table*}

We found that some atoms, such as C, O, V and Pt that are adsorbed
on the H-passivated armchair graphane NR surface show a tendency to
remove H atoms from graphane surface. Since the single atoms of C,
O, V and Pt have rather high binding energy, it seems possible to
create graphene domains on graphane by striping H atoms but covering
the domain by adsorbed atoms. Independent of their initial position,
C and O atoms are adsorbed to carbon atoms of graphane by replacing
H atoms. A typical binding configuration of C and O is indicated as
M1 in Fig.~\ref{adatom}. Adsorption of C and O atoms occur at the
middle of nanoribbon with strong binding energies of 5.89 and 3.99
eV, respectively. The resulting system is NM. Binding structures of
Pt and V, which also remove H atom from the host C atom of graphane,
are also shown in the same figure.

Although H has very low binding energy (and hence is not shown in Table~\ref{tab:adsorp}), it may be important for striping H atoms from graphane. Upon all
geometry relaxations from different initial conditions, single
adsorbed H atom prefers to bind on top of H atom of graphane and
forms H$_2$ molecule which is weakly bound to the nearest C atom.
The binding energy of H$_2$ molecule is calculated to be $\sim 36$
meV.

\section{Vacancies in Graphane NRs}\label{Vacancies}

In our recent study\cite{hasan-can}, we showed that a graphane,
which has a NM ground state, can be made magnetic simply by removing
hydrogen atoms from the uniform hydrogen layers adsorbed on its both
sides. It was found that the magnetization depends on whether the
defect region is one-sided or two-sided. It was also shown that in
certain circumstances remarkably large magnetic moments can be
attained in a small domain on the sheet of graphane. Experimentally,
removal of surface hydrogen atoms by using laser beam resonating
with surface-atom bond\cite{laser}, ionic vapor
application\cite{breaux} and subnanometer Pt clusters\cite{vajda}
are reported before. Since it is revealed that the double-sided
vacancies give higher magnetic moments we focus on these type of
vacancies on the nanoribbons.

\begin{figure}
\includegraphics[width=8.5cm]{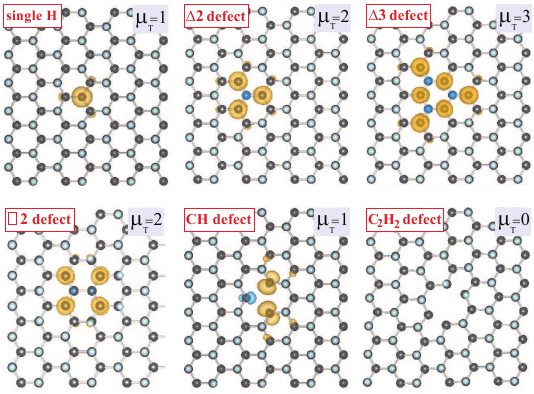}
\caption{(Color online)  Atomic structures corresponding to
single-H, double-sided triangular shaped $\Delta$2 and $\Delta$3,
double-sided rectangular shaped $\square$2, CH and C$_{2}$H$_{2}$
vacancies and resulting difference charge density, $\Delta
\rho=\rho^{(\uparrow)}-\rho^{(\downarrow)}$, surfaces for a bare
armchair graphane NR. Large/black and small/light blue-gray balls
indicate C and H atoms, respectively. Only a small part which
includes vacancy region and its nearby atoms, of the armchair
graphane NR with $N$=15 is shown.} \label{vacancy}
\end{figure}

In this section we consider six different types of vacancies in a
bare armchair graphane NR with $N$=15. Upon desorption of a single
hydrogen atom, as shown in Fig.~\ref{vacancy}, local bonding through
$sp^{3}$-hybrid orbital is retransformed into planar $sp^{2}$- and
perpendicular $p_{z}$-orbitals. At the vacancy site one electron
accommodated by the dangling $p_{z}$-orbital becomes unpaired and
hence contributes to the magnetization by 1 $\mu_{B}$. We also
calculated that a single H-vacancy located at the center of the NR
has 70 meV lower energy than that located near the edge of the
ribbon.

The situation is rather intriguing for the magnetization of large
domains of hydrogen vacancies. As in the case of single H-vacancy,
leading to transformation of $sp^{3}$-bonding into planar
$sp^{2}$-bonding, triangular $\Delta$, and rectangular $\square$
shaped large graphene islands can be created by removing H atoms
hence by creating H-vacancy domains. As shown in Fig.~\ref{vacancy},
hydrogen desorbed triangle-shaped island consisting of $n$ carbon
atom at its each edge is defined as $\Delta$n-type H-vacancy.
Therefore, a $\Delta$n-type vacancy domain can be obtained by
removing $n(n+1)/2$ H atom from top side  and $n(n-1)/2$ H atom from
down side  of the graphane nanoribbon. This means that $n(n+1)/2$
$p_{z}$-orbital electrons freed from A-type and $n(n-1)/2$
$p_{z}$-orbital electrons freed from B-type sublattice C atoms.
According to the Lieb's theorem \cite{Lieb} net magnetic moment of
such a system can be calculated from the difference between the
sublattice atoms. For triangle vacancies this difference is simply
equal to $n$ and thus the $\Delta$n-vacancy domain has net magnetic
moment with $n\mu_{B}$. However, the $\square$-type domains does not
yield large magnetic moments and have values between 0 and 2
$\mu_{B}$. It is seen that magnetic interactions in double-sided
triangular and square H-vacancy domains resulting in net magnetic
moments are relatively straightforward and are in good agreement
with Lieb's theorem.

We also consider CH- and C$_{2}$H$_{2}$-vacancies, the synthesis of
them are relatively easier than creating only H-vacancies. In our
study, these vacancies are first created in a NR by removing
involved atoms and subsequently their structures are optimized.
Calculated values of magnetic moments of the structure are presented
in Fig.~\ref{vacancy}. After the creation of a CH-vacancy on a
graphane nanoribbon, geometric structure rearranges itself via
Jahn-Teller-like distortion in the lattice. Charge density plot in
Fig.~\ref{vacancy} shows the unpaired up and down spin electron
states located on the region surrounding the vacancy. It is seen
that a bare armchair graphane NR, which is NM semiconductor when it
is defect free, attains permanent magnetic moment in the presence of
CH-vacancies. While CH-type vacancy makes small changes in the
geometric structure, Stone-Wales type deformation occurs after
formation of the C$_{2}$H$_{2}$-vacancy. Since electrons are paired
in C$_{2}$H$_{2}$-vacancy, the ground state is again NM and thus
there is no visible difference charge density in Fig.~\ref{vacancy}.
From the point view of Lieb's theorem, removal of a CH couple
results in an absence in the number of $p_{z}$-orbital electron
belonging to a A-type (or B-type) sublattice. Upon the creation of
C$_{2}$H$_{2}$-vacancy, both the A- and B-type sublattices lose one
electron and thus the difference is zero resulting in a NM ground
state. Our results are consistent with the recently reported values
for CH-vacancies in infinite sheets of graphane.\cite{ch-vacancy}

\section{Edge Roughness}\label{Roughness}

\begin{figure}
\includegraphics[width=8.5cm]{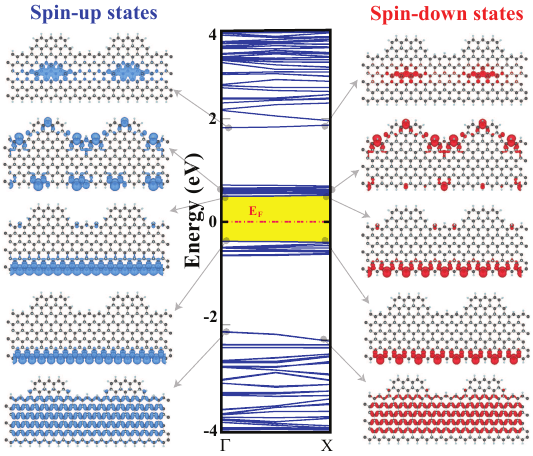}
\caption{(Color online) Energy band diagram and band projected
charge density isosurfaces of various states for bare zigzag
graphane NR including edge roughness. The band gap between edge
states are shaded yellow/gray. Zero of the band energy is set to the
Fermi level.} \label{roughness}
\end{figure}

Earlier, it was shown that the edge roughness of graphene NRs can
affect their electronic\cite{Gunlycke, Yoon, Evaldsson} and
mechanical\cite{can-mechanical} properties. In particular, it was
revealed that periodically repeating edge profiles can be treated
within the superlattice structure, which result in confinement of
spin states in zigzag NRs.\cite{haldun-super} Band alignments of
these superlattice structures have been studied both experimentally
and theoretically.\cite{dai} Here we investigated the effect of
periodically repeating edge roughness of a bare zigzag graphane NR.
While one edge of NR is kept flat, periodically repeating
undulations are carved at the other edge. The same structure can be
viewed as periodically repeating heterostructures of wide ($N$=8)
and narrow ($N$=6) segments of zigzag graphane NRs. Atomic and
electronic band structure of this superlattice, which mimics the
edge roughness are shown in Fig.~\ref{roughness}. Here we note that
infinite bare zigzag graphane NR with $N$=8 has a band gap of 0.8
eV, while the band gap of infinite bare zigzag graphane NR with
$N$=6 is relatively wider and 0.9 eV. As pointed out in Sec. IV,
wide band gap of H-passivated graphane NRs is reduced to 0.8-0.9 eV
because of edge states of unpassivated dangling bonds of bare
graphane NRs, which appear in the band gap. These edge states in
Fig.~\ref{roughness} of the superlattice occur as several closely
lying flat bands below and above the Fermi level separated  by a
superlattice band gap of $\sim$0.8 eV. The character of various
bands are revealed by plotting the charge density isosurfaces of
various states. Charge density isosurfaces of the highest filled
spin up and spin down edge state indicate the localization at the flat
edge. As for the charge distribution of the lower lying valence
states is rather uniform along the narrow region. The states of the
flat band at the edge of the conduction band is confined in the wide
segment of the superlattice since they cannot find a matching state
in the narrow region. As a proof of the concept, it is shown that
electronic states can be confined at specific regions of periodic
edge roughness. Of course, confinement, superlattice band gap etc.
depend on the structural parameters of superlattices and need
detailed investigation.

\section{Discussions and Conclusions}\label{Discussion}

Research on recently synthesized graphane revealed interesting
electronic and magnetic properties of this two dimensional
honeycomb structure. Graphane attains a $sp^3$-like bonding
through covering of its both sides by hydrogen atom. This bonding
is rather different from the $sp^2$-bonding of graphene and
attributes a number of additional properties to graphane. For
example, in contrast to semimetallic graphene with linear band
crossing at the Fermi level, graphane is a nonmagnetic, wide band
semiconductor.

The armchair and zigzag graphane NRs of graphane display important
number of properties and hence constitute basic structures to
fabricate various devices. Both zigzag and armchair graphane NRs are
wide band gap semiconductor when their edges are passivated with
hydrogen. The band gaps vary exponentially with their widths. For
narrow graphane NRs the band gap is large due to quantum confinement
effect, but approaches to the band gap of graphane as the width
increases. Unpaired, unpassivated dangling bonds at the edges have 1
Bohr magneton magnetic moment and have antiferromagnetic coupling
with adjacent dangling bonds. These graphane NRs can be
functionalized to attain additional properties through H-passivation
of their edges, adatom adsorption, vacancy creation, edge profiling
and superlattice formation. In particular, graphane NRs attain
magnetic moment through the creation of H vacancy at the surfaces of
graphane. This property can be utilized to achieve interesting
functionalities through decoration and patterning of H-vacancies on
the graphane NRs. The possibility of generation of large magnetic
moments at small domains of H-vacancies makes graphane based
structures promising for data storage and nanospintronic
applications. These functionalities can be further extended by
adsorbing adatoms to carbon atoms deprived from hydrogen.

In conclusion, the present study demonstrates that graphane
NRs can be an important basic nanomaterial and presents interesting properties for future technological applications.

\begin{acknowledgments}
Computing resources used in this work were partly provided by the
National Center for High Performance Computing of Turkey (UYBHM)
under grant number 2-024-2007. This work is partially supported by
the project of The State Planning Organization (DPT) of Turkey and
by Academy of Science of Turkey (T\"{U}BA).
\end{acknowledgments}

\end{document}